\documentclass[aps,prd,amsmath,superscriptaddress,nofootinbib,%
               twocolumn]{revtex4}

\usepackage{graphicx}

\newcommand{\ie}{\textit{i.e.}}
\newcommand{\eg}{\textit{e.g.}}

\newcommand{\rmd}{\ensuremath{\mathrm{d}}}
\newcommand{\orderof}[1]{\ensuremath{\mathcal{O}(#1)}}

\newcommand{\gae}{%
  \ensuremath{\lower 2pt \hbox{%
    $\, \buildrel {\scriptstyle >}\over {\scriptstyle \sim}\,$}%
    }%
  }
\newcommand{\lae}{%
  \ensuremath{\lower 2pt \hbox{%
    $\, \buildrel {\scriptstyle <}\over {\scriptstyle \sim}\,$}%
    }%
  }

\newcommand{\mpl}{\ensuremath{m_\textrm{Pl}}}
\newcommand{\mgut}{\ensuremath{m_\textrm{GUT}}}
\newcommand{\phie}{\ensuremath{\phi_\textrm{e}}}
\newcommand{\Ve}{\ensuremath{V_\textrm{e}}}
\newcommand{\Vp}{\ensuremath{V^{\prime}}}
\newcommand{\Vpp}{\ensuremath{V^{\prime\prime}}}
\newcommand{\ns}{\ensuremath{n_s}}
\newcommand{\nt}{\ensuremath{n_\textrm{T}}}
\newcommand{\nsrun}{%
  \ensuremath{\frac{\mathrm{d} n_s}{\mathrm{d} \ln{k}}}%
  }
\newcommand{\Pscalar}{\ensuremath{P_\mathcal{R}}}
\newcommand{\Pscalarrt}{\ensuremath{\Pscalar^{1/2}}}
\newcommand{\Ptensor}{\ensuremath{P_T}}
\newcommand{\Ptensorrt}{\ensuremath{\Ptensor^{1/2}}}
\newcommand{\Pratio}{\ensuremath{\frac{P_T}{P_\mathcal{R}}}}
\newcommand{\rhoRH}{\ensuremath{\rho_\textrm{RH}}}

\newcommand{\refeqn}[2][eqn:]{Eqn.~(\ref{#1#2})}

\newcommand{\reffig}[2][fig:]{Figure~\ref{#1#2}}
\newcommand{\refFig}[2][fig:]{Figure~\ref{#1#2}}
\newcommand{\refsec}[2][sec:]{Section~\ref{#1#2}} 

\newcommand{\insertfig}[1]{%
    \includegraphics[keepaspectratio,width=1.00\columnwidth,
                     height=0.40\textheight]{#1}
}

\begin{document}


\preprint{MCTP-06-23}
\preprint{FTPI-MINN-06/32}


\title{Natural Inflation: status after WMAP 3-year data}

\author{Christopher Savage}
\email[]{cmsavage@umich.edu}
\altaffiliation{
 Also at
 William I. Fine Theoretical Physics Institute,
 School of Physics \& Astronomy,
 University of Minnesota,
 Minneapolis, MN 55455}
\affiliation{
 Michigan Center for Theoretical Physics,
 Department of Physics,
 University of Michigan,
 Ann Arbor, MI 48109}

\author{Katherine Freese}
\email[]{ktfreese@umich.edu}
\affiliation{
 Michigan Center for Theoretical Physics,
 Department of Physics,
 University of Michigan,
 Ann Arbor, MI 48109}

\author{William H. Kinney}
\email[]{whkinney@buffalo.edu}
\affiliation{
 Department of Physics,
 Univerity at Buffalo, SUNY,
 Buffalo, NY 14260}

\date{\today}


\begin{abstract}

The model of Natural Inflation is examined in light of recent 3-year
data from the Wilkinson Microwave Anisotropy Probe and shown to
provide a good fit.  The inflaton potential is naturally flat due to
shift symmetries, and in the simplest version takes the form $V(\phi)
= \Lambda^4 [1 \pm \cos(N\phi/f)]$.  The model agrees with
WMAP3 measurements as long as $f > 0.7 \mpl$ (where
$\mpl = 1.22 \times 10^{19}$GeV) and $\Lambda \sim \mgut$.  The
running of the scalar spectral index is shown to be small -- an order
of magnitude below the sensitivity of WMAP3. The location of the field
in the potential when perturbations on observable scales are produced
is examined; for $f>5\mpl$, the relevant part of the  potential is
indistinguishable from a quadratic, yet has the advantage that the
required flatness is well-motivated.  Depending on the value of $f$,
the model falls into the large field ($f \ge 1.5 \mpl$) or small field
($f<1.5 \mpl$) classification scheme that has been applied to inflation
models. Natural inflation provides a good fit to WMAP3 data.

\end{abstract}

\maketitle


\section{\label{sec:Intro} Introduction}

\begin{figure}
  \insertfig{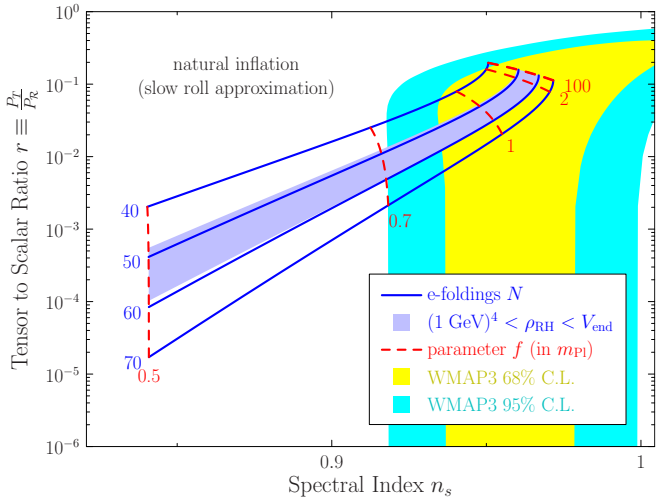}
  \caption[$r-n_s$ plane] {
    Natural inflation predictions and WMAP3 constraints in the $r$-$\ns$
    plane.  (Solid/blue) lines running from approximately the lower left
    to upper right are predictions for constant $N$ and varying $f$,
    where $N$ is the number of e-foldings prior to the end of inflation
    at which current modes of scale $k = 0.002$ Mpc$^{-1}$
    were generated and $f$ is the width of the potential.
    The remaining (dashed/red) lines are for constant $f$ and varying
    $N$.
    The (light blue) band corresponds to the values of $N$ for
    standard post-inflation cosmology with $(\textrm{1 GeV})^4 <
    \rhoRH < V_\textrm{end}$.  Filled (nearly vertical) regions are the
    parameter spaces allowed by WMAP3 at 68\% and 95\% C.L.'s
    (error contours taken from Ref.~\cite{Kinney:2006qm}).
    Natural inflation is consistent with the WMAP3 data for
    $f \gae 0.7\mpl$ and essentially all likely values of $N$.
    }
  \label{fig:rnplane}
\end{figure}

Inflation was proposed \cite{Guth:1980zm,Kazanas:1980tx,
Starobinsky:1980te,Sato:1981ds,Sato:1980yn} to solve several
cosmological puzzles: an early period of accelerated expansion explains
the homogeneity, isotropy, and flatness of the universe, as well as the
lack of relic monopoles.  While inflation results in an approximately
homogeneous universe, inflation models also predict small
inhomogeneities.  Observations of inhomogeneities via the cosmic
microwave background (CMB) anisotropies and structure formation are now
providing tests of inflation models.

The release of three years of data from the Wilkinson Microwave
Anisotropy Probe (WMAP3) \cite{Spergel:2006hy} satellite have generated
a great deal of excitement.  First, generic predictions of inflation
match the observations: the universe has a critical density
($\Omega=1$), the density perturbation spectrum is nearly scale
invariant, and superhorizon fluctuations are evident. Second, current
data is beginning to differentiate between inflationary models and
already rules some of them out \cite{Spergel:2006hy,Alabidi:2006qa,
Peiris:2006ug,Easther:2006tv,Seljak:2006bg,Kinney:2006qm,Martin:2006rs,
Peiris:2006sj}. (For example, quartic potentials  and generic tree-level
hybrid models do not provide a good match to the data.)
It is the purpose of this paper to illustrate that the model
known as Natural Inflation is an excellent  match to current data.

Inflation models predict two types of perturbations, scalar and tensor,
which result in density and gravitational wave fluctuations,
respectively.  Each is typically characterized by a fluctuation
amplitude ($\Pscalarrt$ for scalar and $\Ptensorrt$ for tensor, with
the latter usually given in terms of the ratio $r \equiv
\Ptensor/\Pscalar$) and a spectral index ($\ns$ for scalar and
$\nt$ for tensor) describing the mild scale dependence of the
fluctuation amplitude.  The amplitude $\Pscalarrt$ is normalized
by the height of the inflationary potential.  The inflationary
consistency condition $r = -8 \nt$ further reduces the number of free
parameters to two, leaving experimental limits on $\ns$ and $r$ as the
primary means of distinguishing among inflation models.  Hence,
predictions of models are presented as plots in the $r$-$\ns$ plane.

Most inflation models suffer from a potential drawback: to match
various observational constraints, notably CMB anisotropy measurements
as well as the requirement of sufficient inflation,
the height of the inflaton potential must be of a much smaller scale
than that of the width, by many orders of magnitude (\ie, the potential
must be very flat).  This requirement of two very different mass scales
is what is  known as the ``fine-tuning'' problem in inflation, since
very precise couplings are required in the theory to prevent radiative
corrections from bringing the two mass scales back to the same level.
The natural inflation model (NI) uses shift symmetries to generate a
flat potential, protected from radiative corrections, in a natural way
\cite{Freese:1990rb}.  In this regard, NI is one of the best motivated
inflation models.  

One of the major results of the paper is shown in \reffig{rnplane}.
The predictions of NI are plotted in the $r$-$\ns$ plane for various
parameters: the width $f$ of the potential and number of e-foldings $N$
before the end of inflation at which present day fluctuation modes of
scale $k=0.002$ Mpc$^{-1}$ were produced.  $N$ depends upon the
post-inflationary universe and is $\sim$50-60.  Also shown in the
figure are the observational constraints from WMAP's recent 3-year
data, which provides some of the tightest constraints on inflationary
models to date \cite{Spergel:2006hy}.  The primary result is that NI,
for $f \gae 0.7\mpl$, is consistent with current observational
constraints.

In this paper we take $\mpl = 1.22 \times 10^{19}$ GeV.  Our result
extends upon a previous analysis of NI \cite{Freese:2004un}
that was based upon WMAP's first year data \cite{Spergel:2003cb}.
Earlier analyses \cite{Adams:1992bn,Moroi:2000jr} have placed
observational constraints on this model using COBE data
\cite{Smoot:1992td}.  Other papers have more recently considered NI in
light of the WMAP1 and WMAP3 data \cite{Alabidi:2005qi,Alabidi:2006qa}.

This paper emphasizes two further results as well. First, we
investigate the running of the spectral index in natural inflation,
\ie\ the dependence of $\ns$ on scale, and find that it is small: two
orders of magnitude smaller than the sensitivity of WMAP3 and below the
sensitivity of any planned experiment.  Second, we
find how far down the potential the field is at the time structure is
produced, and find that for $f > 5 \mpl$ the relevant part of the
potential is indistinguishable from a quadratic potential.  (Still,
the naturalness motivation for NI renders it a superior model to a
quadratic potential as the latter typically lacks 
an explanation for its flatness).

We will begin by discussing the model of natural inflation in
\refsec{NI}: the motivation, the potential, the evolution of the
inflaton field, and relating pre- and post-inflation scales.  In
 \refsec{Fluctuations}, we will examine the scalar and tensor
perturbations predicted by NI and compare them with the WMAP 3-year
data. In \refsec{Running}, we will address the running of the spectral
index.  In \refsec{Potential}, we will examine the location on the
potential at which the observable e-folds of inflation take place and
examine where NI falls in the small field/large field/hybrid model
categorization scheme.  We conclude in \refsec{Conclusion}.

\section{The Model of Natural Inflation\label{sec:NI}}


\textit{Motivation:}
To satisfy a combination of constraints on inflationary models, in
particular, sufficient inflation and microwave background anisotropy
measurements \cite{Spergel:2003cb,Spergel:2006hy}, the
potential for the inflaton field must be very flat.  For a general
class of inflation models involving a single slowly-rolling field, it
has been shown that the ratio of the height to the (width)$^4$ of the
potential must satisfy \cite{Adams:1990pn}
\begin{equation} \label{eqn:Vratio}
  \chi \equiv \Delta V/(\Delta \phi)^4 \le {\cal O}(10^{-6} - 10^{-8})
  \, , 
\end{equation}
where $\Delta V$ is the change in the potential $V(\phi)$ and $\Delta
\phi$ is the change in the field $\phi$ during the slowly rolling
portion of the inflationary epoch.  Thus, the inflaton must be
extremely weakly self-coupled, with effective quartic self-coupling
constant $\lambda_{\phi} < \orderof{\chi}$ (in realistic models,
$\lambda_{\phi} < 10^{-12}$).  The small ratio of mass scales required
by \refeqn{Vratio} quantifies how flat the inflaton potential must be
and is known as the ``fine-tuning'' problem in inflation.
A recent review of inflation can be found in Ref.~\cite{Bassett:2005xm}.

Three approaches have been taken toward this required flat potential
characterized by a small ratio of mass scales.  First, some simply say
that there are many as yet unexplained hierarchies in physics, and
inflation requires another one.  The hope is that all these
hierarchies will someday be explained.  In these cases, the tiny
coupling $\lambda_{\phi}$ is simply postulated \textit{ad hoc} at tree
level, and then must be fine-tuned to remain small in the presence of
radiative corrections.  But this merely replaces a cosmological
naturalness problem with unnatural particle physics.  Second, models
have been attempted where the smallness of $\lambda_{\phi}$ is
protected by a symmetry, \eg, supersymmetry.  In these cases (\eg,
\cite{Holman:1984yj}), $\lambda_{\phi}$ may arise from a small ratio
of mass scales; however, the required mass hierarchy, while stable, is
itself unexplained.  In addition, existing models have limitations.
It would be preferable if such a hierarchy, and thus inflation itself,
arose dynamically in particle physics models.

Hence, in 1990 a third approach was proposed, Natural Inflation
\cite{Freese:1990rb}, in which the inflaton potential is flat due to
shift symmetries.  Nambu-Goldstone bosons (NGB) arise whenever a
global symmetry is spontaneously broken.  Their potential is exactly
flat due to a shift symmetry under $\phi \rightarrow \phi + \textrm{
constant}$. As long as the shift symmetry is exact, the inflaton
cannot roll and drive inflation, and hence there must be additional
explicit symmetry breaking.  Then these particles become pseudo-Nambu
Goldstone bosons (PNGBs), with ``nearly'' flat potentials, exactly as
required by inflation.  The small ratio of mass scales required by
\refeqn{Vratio} can easily be accommodated. For example, in the case
of the QCD axion, this ratio is of order $10^{-64}$.  While inflation
clearly requires different mass scales than the axion, the point is
that the physics of PNGBs can easily accommodate the required small
numbers.

The NI model was first proposed and a simple analysis performed in
\cite{Freese:1990rb}.  Then, in 1993, a second paper followed which
provides a much more detailed study \cite{Adams:1992bn}.  
Many types of candidates have subsequently been explored for natural
inflation.  For example, WHK and K.T.\ Mahanthappa considered NI
potentials generated by radiative corrections in models with explicitly
broken Abelian \cite{Kinney:1995xv} and non-abelian \cite{Kinney:1995cc}
symmetries, showing that NI models with $f \sim \mpl$ and $f \ll \mpl$
can both be generated in self-consistent field theories.
Ref.~\cite{Kawasaki:2000yn} used shift symmetries
in Kahler potentials to obtain a flat potential and drive natural
chaotic inflation in supergravity.  Additionally,
\cite{Arkani-Hamed:2003wu,Arkani-Hamed:2003mz} examined natural
inflation in the context of extra dimensions and \cite{Kaplan:2003aj}
used PNGBs from little Higgs models to drive hybrid inflation.  Also,
\cite{Firouzjahi:2003zy,Hsu:2004hi} use the natural inflation idea of
PNGBs in the context of braneworld scenarios to drive inflation.
Freese \cite{Freese:1994fp} suggested using a PNGB as the rolling
field in double field inflation \cite{Adams:1991ma} (in which the
inflaton is a tunneling field whose nucleation rate is controlled by
its coupling to a rolling field).  We will focus in this paper on the
original version of natural inflation, in which there is a single
rolling field.


\textit{Potential:}
The PNGB potential resulting from explicit breaking of a shift symmetry
in single field models (in four spacetime dimensions) is generally of
the form
\begin{equation} \label{eqn:potential}
  V(\phi) = \Lambda^4 [1 \pm \cos(N\phi/f)] \, .
\end{equation}
We will take the positive sign in \refeqn{potential} (this choice has
no effect on our results) and take $N = 1$, so the potential, of
height $2 \Lambda^4$, has a unique minimum at $\phi = \pi f$ (the
periodicity of $\phi$ is $2 \pi f$).

For appropriately chosen values of the mass scales, \eg\ $f \sim \mpl$
and $\Lambda \sim \mgut \sim 10^{15}$ GeV, the PNGB field $\phi$ can
drive inflation.  This choice of parameters indeed produces the small
ratio of scales required by \refeqn{Vratio}, with $\chi \sim
(\Lambda/f)^4 \sim 10^{-13}$.  While $f \sim \mpl$ seems to be a
reasonable scale for the potential width, there is no reason to
believe that $f$ cannot be much larger than $\mpl$.  In fact, Kim,
Nilles \& Peloso \cite{Kim:2004rp} as well as the idea of N-flation
\cite{Dimopoulos:2005ac} showed that an \textit{effective} potential of
$f \gg \mpl$ can be generated from two or more axions, each with
sub-Plankian scales.  We shall thus include the possibility of
$f \gg \mpl$ is our analysis and show that these parameters can fit the
data.

\begin{figure}
  \insertfig{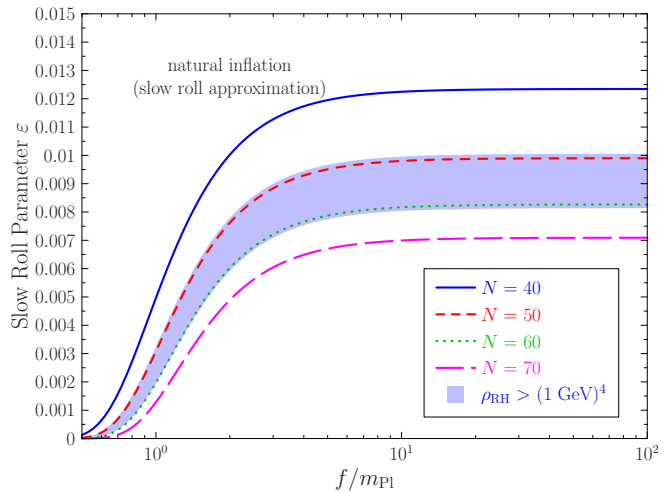}
  \caption[Slow roll parameter $\epsilon$]{
    The slow roll parameter $\epsilon$ is shown as a function of the
    potential width $f$ for various numbers of e-foldings $N$ before the
    end of inflation.
    The (light blue) band corresponds to the values
    of $N$ consistent with the standard post-inflation cosmology, as
    given by \refeqn{Nk}, for an end of reheating energy density
    $(\textrm{1 GeV})^4 < \rhoRH < V_\textrm{end}$, where the lower
    bound is a result of nucleosynthesis constraints.
    }
  \label{fig:epsilon}
\end{figure}

\textit{Evolution of the Inflaton Field:}
The evolution of the inflaton field is described by
\begin{equation} \label{eqn:eom}
  \ddot{\phi} + 3H\dot{\phi} + \Gamma\dot{\phi} + \Vp(\phi) = 0
  \, ,
\end{equation}
where $\Gamma$ is the decay width of the inflaton.  A sufficient
condition for inflation is the slow-roll (SR) condition $\ddot{\phi} \ll
3 H \dot{\phi}$.  The expansion of the scale factor $a$, with $H =
\dot{a}/a$, is determined by the scalar field dominated Friedmann
equation,
\begin{equation} \label{eqn:friedman}
  H^2 = \frac{8\pi}{3\mpl^2} V(\phi) .
\end{equation}
The slow roll (SR) condition implies that two conditions are met:
\begin{eqnarray} \label{eqn:epsilonA}
  \epsilon(\phi)
    &\approx& \frac{\mpl^2}{16\pi}
              \left[ \frac{\Vp(\phi)}{V(\phi)} \right]^2
              \nonumber\\
    &=&       \frac{1}{16\pi}
              \left( \frac{\mpl}{f} \right)^2
              \left[ \frac{\sin(\phi/f)}{1+\cos(\phi/f)} \right]^2 \ll 1
\end{eqnarray}
and
\begin{eqnarray} \label{eqn:etaA}
  \eta(\phi)
    &\approx& \frac{\mpl^2}{8\pi}
              \left[ \frac{\Vpp(\phi)}{V(\phi)}
                      - \frac{1}{2} \left(
                          \frac{\Vp(\phi)}{V(\phi)} \right)^2
              \right]  \nonumber\\
    &=&       - \frac{1}{16\pi} \left( \frac{\mpl}{f} \right)^2 \, \ll 1.
\end{eqnarray}
Inflation ends when the field $\phi$ reaches a value $\phie$ such that
$\epsilon(\phi) < 1$ is violated, or
\begin{equation} \label{eqn:phie}
  \cos(\phie/f) = \frac{1 - 16\pi(f/\mpl)^2}{1 + 16\pi(f/\mpl)^2} \, .
\end{equation}
\refFig{epsilon} illustrates the value of $\epsilon$ during periods
where density fluctuations are produced; one can see that indeed
$\epsilon \ll 1$.

More accurate results can be attained by numerically solving the
equation of motion, \refeqn{eom}, together with the Friedmann
equations.  Such calculations have been performed in
Ref.~\cite{Adams:1992bn}, where it was shown the SR analysis is
accurate to within a few percent for the $f \gae 0.5\mpl$ parameter
space we will be examining.  Thus, we are justified in using the SR
approximation in our calculations.


\textit{Relating Pre- and Post-Inflation Scales:}
To test inflationary theories, present day observations must be
related to the evolution of the inflaton field during the inflationary
epoch.  Here we show how a comoving scale $k$ today can be related
back to a point during inflation.  We need to find the value of $N_k$,
the number of e-foldings before the end of inflation, at which
structures on scale $k$ were produced.

Under a standard post-inflation cosmology, once inflation ends, the
universe undergoes a period of reheating. Reheating can be
instantaneous or last for a prolonged period of matter-dominated
expansion.  Then reheating ends at $T <T_\textrm{RH}$, and the
universe enters its usual radiation-dominated and subsequent
matter-dominated history.  Instantaneous reheating ($\rhoRH = \rho_e$)
gives the minimum number of e-folds as one looks backwards
to the time of perturbation production, while a prolonged period of
reheating gives a larger number of e-folds.

The relationship between scale $k$ and the number of e-folds $N_k$
before the end of inflation has been shown to be \cite{Lidsey:1995np}
\begin{equation} \label{eqn:Nk}
  N_k = 62 - \ln\frac{k}{a_0 H_0}
            - \ln\frac{10^{16}\,\textrm{GeV}}{V_k^{1/4}}
            + \ln\frac{V_k^{1/4}}{\Ve^{1/4}}
            - \frac{1}{3} \ln\frac{\Ve^{1/4}}{\rhoRH^{1/4}} \, .
\end{equation}
Here, $V_k$ is the potential when $k$ leaves the horizon during
inflation, $\Ve = V(\phie)$ is the potential at the end of inflation,
and $\rhoRH$ is the energy density at the end of the reheat period.
Nucleosynthesis generally requires $\rhoRH \gae (\textrm{1 GeV})^4$,
while necessarily $\rhoRH \le \Ve$.  Since $\Ve$ may be of order
$\mgut \sim 10^{15}$ GeV or even larger, there is a broad
allowed range of $\rhoRH$; this uncertainty in $\rhoRH$ translates
into an uncertainty of 10 e-folds in the value of $N_k$ that
corresponds to any particular scale of measurement today.

Henceforth we will use $N$ to refer to the number of e-foldings prior
to the end of inflation that correspond to scale $k = 0.002$
Mpc$^{-1}$, the scale at which WMAP presents their results%
\footnote{The current horizon scale corresponds to
  $k \approx 0.00033$ Mpc$^{-1}$.  The difference in these two scales
  corresponds to only a small difference in e-foldings of
  $\Delta N \lae 2$: while we shall present parameters evaluated at
  $k = 0.002$ Mpc$^{-1}$, those parameters evaluated at the current
  horizon scale will have essentially the same values (at the few
  percent level).
  }.
Under the standard cosmology, this scale corresponds to
$N\sim$50-60 (smaller $N$ corresponds to smaller $\rhoRH$), with a
slight dependence on $f$.  However, if one were to consider
non-standard cosmologies \cite{Liddle:2003as}, the range of possible
$N$ would be broader.  Hence we will show results for the more
conservative range $40 \le N \le 70$, in addition to the more limited
standard cosmology range.

\section{\label{sec:Fluctuations} Perturbations}

As the inflaton rolls down the potential, quantum fluctuations lead
to metric purturbations that are rapidly inflated beyond the horizon.
These fluctuations are frozen until they re-enter the horizon during
the post-inflationary epoch, where they leave their imprint on large
scale structure formation and the cosmic microwave background (CMB)
anisotropy \cite{Guth:1982ec,Hawking:1982cz,Starobinsky:1982ee}.
In this section, we will examine the scalar (density) and tensor
(gravitational wave) purturbations predicted by natural inflation
and compare them with the WMAP 3 year (WMAP3) data
\cite{Spergel:2006hy}.

\subsection{\label{sec:Scalar} Scalar (Density) Fluctuations}

The perturbation amplitude for the density fluctuations (scalar modes) 
produced during inflation is given by
\cite{Mukhanov:1985rz,Mukhanov:1988jd,Mukhanov:1990me,Stewart:1993bc}
\begin{equation} \label{eqn:Pscalar}
   \Pscalarrt(k) = \frac{H^2}{2\pi\dot{\phi}_k} \, .
\end{equation}
Here, $\Pscalarrt(k) \sim \frac{\delta\rho}{\rho}|_\textrm{hor}$ 
denotes the perturbation amplitude when a given wavelength re-enters the
Hubble radius in the radiation- or matter-dominated era, and
the right hand side of \refeqn{Pscalar} is to be evaluated when
the same comoving wavelength ($2\pi/k$) crosses outside the horizon
during inflation.

\begin{figure}
  \insertfig{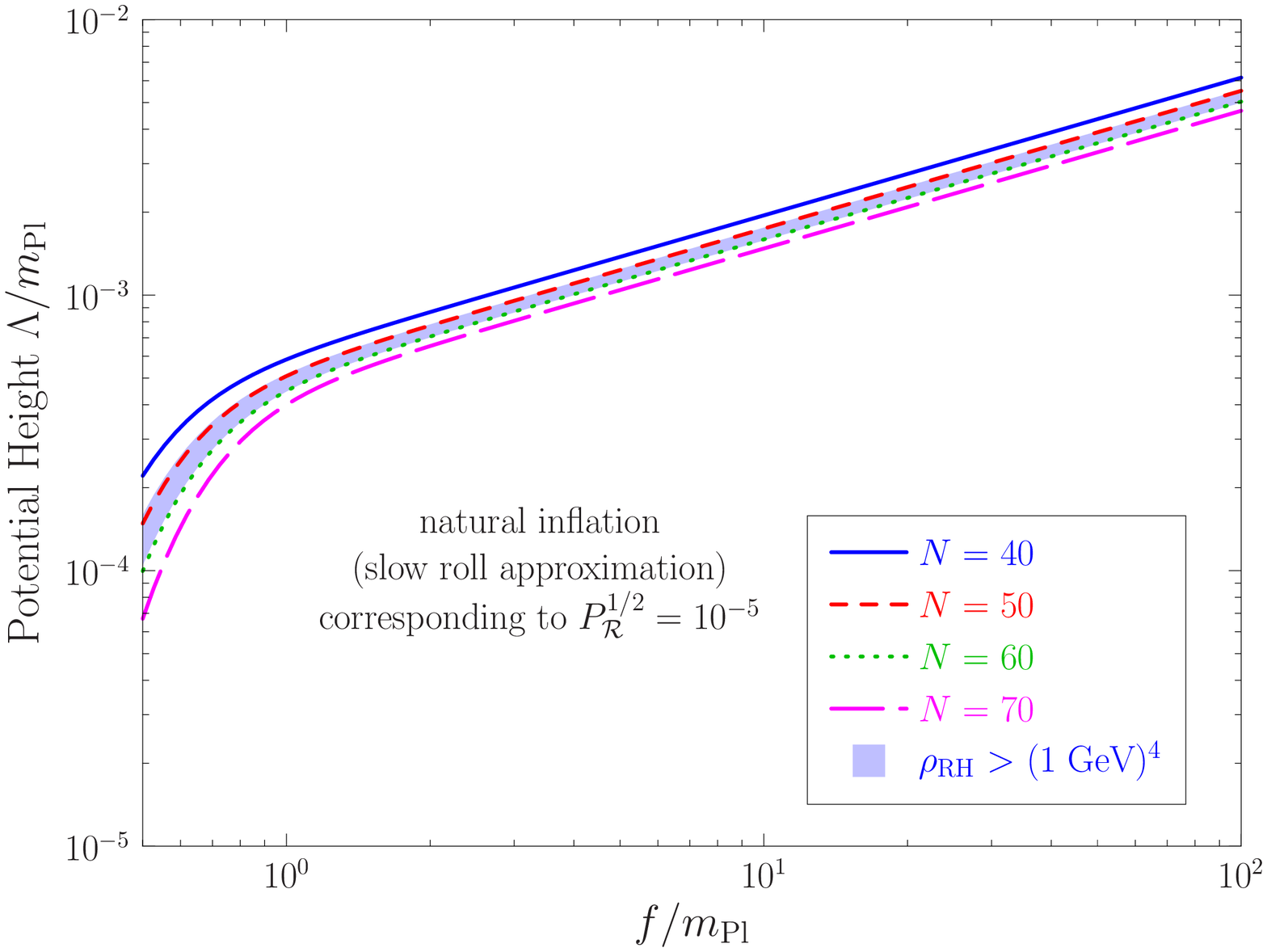}
  \caption[Potential height scale $\Lambda$]{
    The potential height scale $\Lambda$ corresponding to
    $\Pscalarrt = 10^{-5}$ is shown as a function of the potential
    width $f$ for various numbers of e-foldingss $N$ before the end
    of inflation.
    The (light blue) band corresponds to the values of $N$ consistent
    with the standard post-inflation cosmology for
    $\rhoRH > (\textrm{1 GeV})^4$.
    }
  \label{fig:lambda}
\end{figure}

Normalizing to the COBE \cite{Smoot:1992td} or WMAP
\cite{Spergel:2006hy} an\-iso\-tropy measurements gives $\Pscalarrt \sim
10^{-5}$.  This normalization can be used to approximately fix the
height  $\Lambda$ of the potential \refeqn{potential}.  The largest
amplitude perturbations on observable scales are those produced
$N \sim 60$ e-folds before the end of inflation (corresponding to the
horizon scale today), when the field value is $\phi = \phi_N$.  Under
the SR approximation, the amplitude on this scale takes the value
\begin{equation} \label{eqn:Pscalar2}
  \Pscalar \approx \frac{128\pi}{3}
                   \left( \frac{\Lambda}{\mpl} \right)^4
                   \left( \frac{f}{\mpl} \right)^2
                   \frac{[1 + \cos(\phi_N/f)]^3}{\sin^2(\phi_N/f)} \, .
\end{equation}
The values for $\Lambda$ corresponding to $\Pscalarrt = 10^{-5}$ are
shown in \reffig{lambda}.  We see that $\Lambda \sim
10^{15}$-$10^{16}$~GeV for $f \sim \mpl$, yielding an inflaton mass
$m_\phi = \Lambda/f^2 \sim 10^{11}$-$10^{13}$~GeV.  Thus, a potential
height $\Lambda$ of the GUT scale and a potential width $f$ of the
Planck scale are required in NI in order to 
produce the fluctuations responsible for large scale
structure.  For $f \gg \mpl$, the potential height scales as
$\Lambda \sim (10^{-3}\mpl) \sqrt{f/\mpl}$.

\begin{figure}
  \insertfig{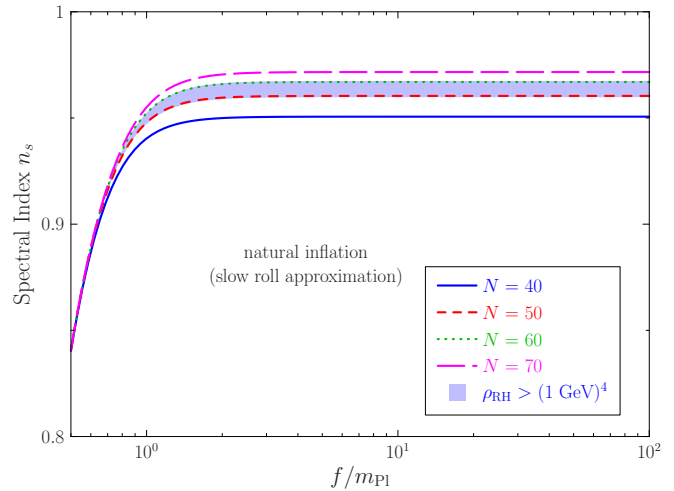}
  \caption[Spectral index $n_s$]{
    The spectral index $n_s$ is shown as a function of the potential
    width $f$ for various numbers of e-foldingss $N$ before the end
    of inflation.
    The (light blue) band corresponds to the values of $N$ consistent
    with the standard post-inflation cosmology for
    $\rhoRH > (\textrm{1 GeV})^4$.
    }
  \label{fig:ns}
\end{figure}

The fluctuation amplitudes are, in general, scale dependent.  The
spectrum of fluctuations is characterized by the spectral index $\ns$,
\begin{equation} \label{eqn:ns}
  \ns - 1 \equiv  \frac{\rmd\ln\Pscalar}{\rmd\ln k}
          \approx -\frac{1}{8\pi} \left( \frac{\mpl}{f} \right)^2
                  \frac{3 - \cos(\phi/f)}{1 + \cos(\phi/f)} \, .
\end{equation}
The spectral index for natural inflation is shown in \reffig{ns}.  For
small $f$, $\ns$ is essentially independent of $N$, while for
$f \gae 2\mpl$, $\ns$ has essentially no $f$ dependence.  Analytical
estimates can be obtained in these two regimes:
\begin{equation} \label{eqn:nsA}
  \ns \approx
  \begin{cases}
    1 - \frac{\mpl^2}{8 \pi f^2} \, ,
      & \textrm{for} \,\, f \lae \frac{3}{4}\mpl \\
    1 - \frac{2}{N} \, ,
      & \textrm{for} \,\, f \gae 2\mpl \, .
  \end{cases}
\end{equation}
Previous analyses of COBE data, based in part on determinations of this
spectral index, have led to constraints on the width of the natural
inflation potential of  $f \gae 0.3\mpl$ \cite{Adams:1992bn} and
$f \gae 0.4\mpl$ \cite{Moroi:2000jr}, while analysis of WMAP's first
year data requires $f \gae 0.6\mpl$ \cite{Freese:2004un}.  Values of $f$
below these constraints would lead to $\ns < 0.9$, reducing fluctuations
at small scales and suppressing higher order acoustic peaks (relative to
lower order peaks) to a level inconsistent with the CMB data.  The WMAP
3-year data yield $\ns = 0.951_{-0.019}^{+0.015}$
($\ns = 0.987_{-0.037}^{+0.019}$ when tensor modes are included in the
fits) on the $k=0.002 {\rm Mpc}^{-1}$ scale%
\footnote{As discussed in \refsec{Running}, the running of the spectral
  index $\ns$ in natural inflation is so small that the value of $\ns$
  at the scale of the WMAP3 measurements
  is virtually identical to its value on the horizon scale.
  }.
This WMAP3 result leads to the somewhat tighter constraint
$f \gae 0.7\mpl$ at 95\% C.L.

\subsection{\label{sec:Tensor} Tensor (Gravitational Wave) Fluctuations}

In addition to scalar (density) perturbations, inflation also produces
tensor (gravitational wave) perturbations with amplitude
\begin{equation} \label{eqn:Ptensor}
  \Ptensorrt(k) = \frac{4H}{\sqrt{\pi}\mpl} \, .
\end{equation}
Here, we examine the tensor mode predictions of natural inflation and
compare with WMAP data.

\begin{figure}
  \insertfig{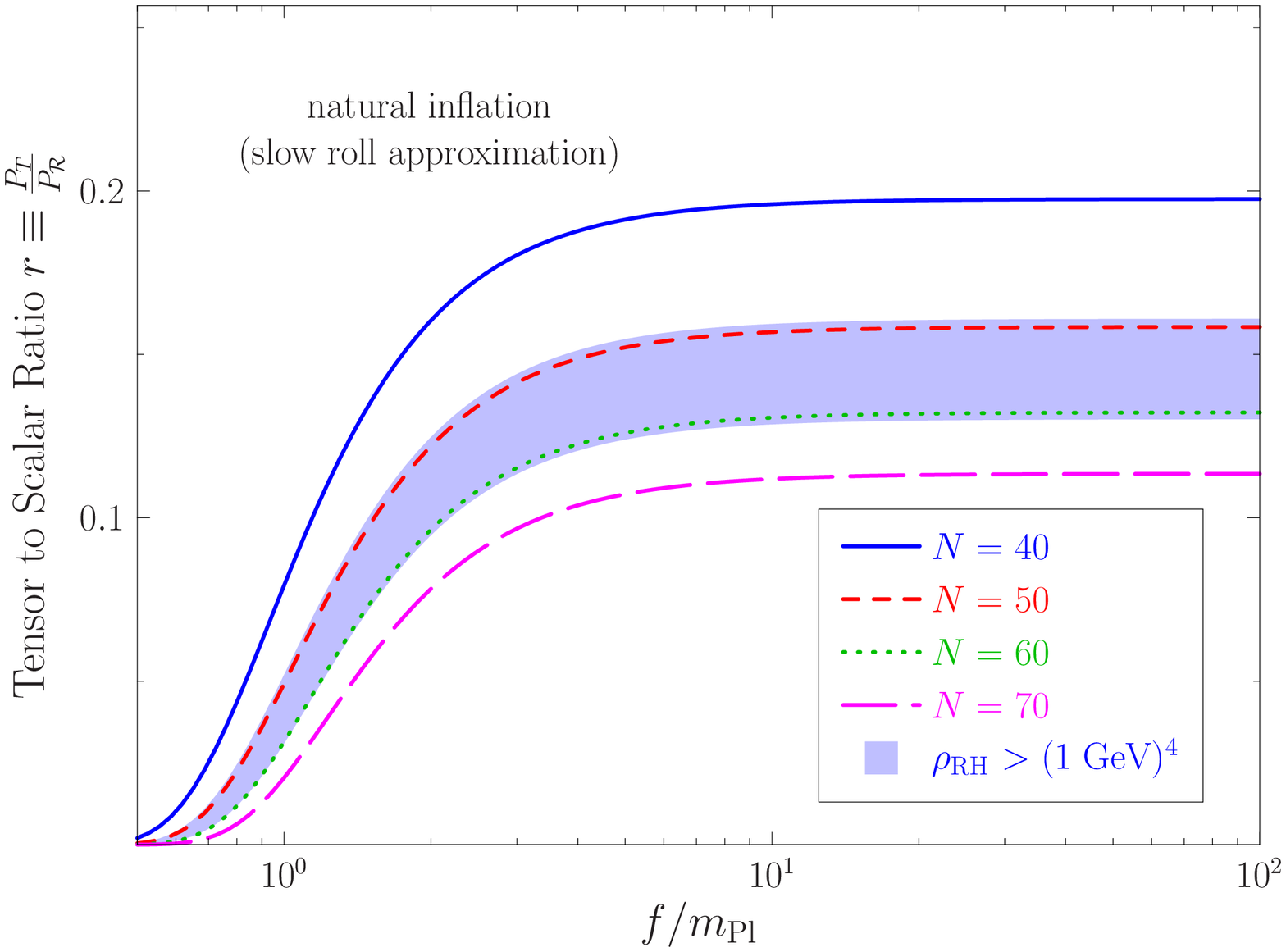}
  \caption[Tensor to scalar ratio $r$]{
    The tensor to scalar ratio $r \equiv \Pratio$ is shown as a function
    of the potential width $f$ for various numbers of e-foldingss $N$
    before the end of inflation.
    The (light blue) band corresponds to the values of $N$ consistent
    with the standard post-inflation cosmology for
    $\rhoRH > (\textrm{1 GeV})^4$.
    }
  \label{fig:ratio}
\end{figure}

Conventionally, the tensor amplitude is given in terms of the
tensor/scalar ratio
\begin{equation} \label{eqn:Pratio}
  r \equiv \Pratio = 16 \epsilon \, ,
\end{equation}
which is shown in \reffig{ratio} for natural inflation.  For small $f$,
$r$ rapidly becomes negligible, while $f \to \frac{8}{N}$ for
$f \gg \mpl$.  In all cases, $r \lae 0.2$, well below the WMAP limit of
$r < 0.55$ (95\% C.L., no running).

As mentioned in the introduction, 
in principle, there are four parameters describing scalar and tensor
fluctuations: the amplitude and spectra of both components, with the
latter characterized by the spectral indices $\ns$ and $\nt$
(we are ignoring any running here).  The amplitude of the scalar
perturbations is normalized by the height of the potential (the energy
density $\Lambda^4$).  The tensor spectral index $\nt$ is not
an independent parameter since it is related to the tensor/scalar ratio
$r$ by the inflationary consistency condition $r = -8 \nt$.
The remaining free parameters are the spectral index $\ns$ of the scalar
density fluctuations, and the tensor amplitude (given by $r$).  

Hence, a useful parameter space for plotting the model predictions
versus observational constraints is on the $r$-$\ns$ plane
\cite{Dodelson:1997hr,Kinney:1998md}.  Natural inflation generically
predicts a tensor amplitude well below the detection sensitivity of
current measurements such as WMAP. However, the situation will improve
markedly in future experiments with greater sensitivity such as QUIET 
\cite{winstein} and PLANCK \cite{unknown:2006uk}, as well as
proposed experiments such as CMBPOL \cite{Bock:2006yf}.

In \reffig{rnplane}, we show the predictions of natural inflation for
various choices of the number of e-folds $N$ and the mass scale $f$,
together with the WMAP3 observational constraints.  Parameters
corresponding to fixed $N=(40,50,60,70)$ with varying $f$ are shown
as (solid/blue) lines from the lower left to upper right.  The
orthogonal (dashed/red) lines correspond to fixed $f$ with varying $N$.
The (blue) band are the values of $N$ consistent with standard
post-inflation cosmology for reheat temperatures above the
nucleosynthesis limit of $\sim$1 GeV, as discussed previously.  The
solid regions are the WMAP3 allowed parameters at 68\% and 95\% C.L.'s.
For a given $N$, a fixed point is reached for $f \gg \mpl$; that is,
$r$ and $\ns$ become essentially independent of $f$ for any
$f \gae 10\mpl$.  This is apparent from the $f=10\mpl$ and $f=100\mpl$
lines in the figure, which are both shown, but are indistinguishable.
As seen in the figure, $f \lae 0.7\mpl$ is excluded.  However,
$f \gae 0.8\mpl$ falls well into the WMAP3 allowed region and is thus
consistent with the WMAP3 data.

\section{\label{sec:Running} Running of the Spectral Index}

\begin{figure}
  \insertfig{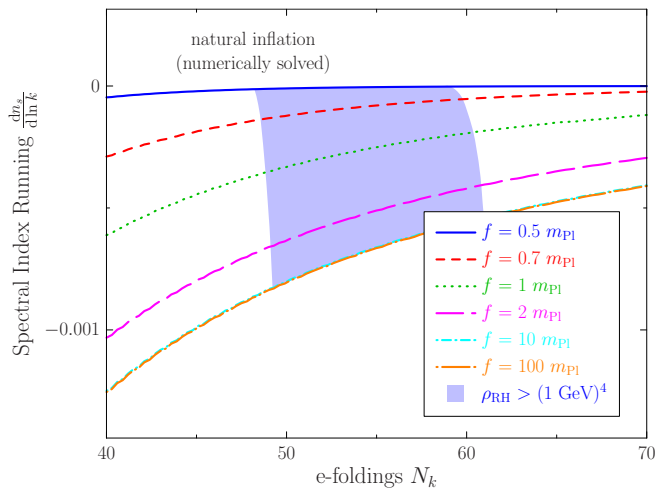}
  \caption[Spectral index running $\nsrun$]{
    The spectral index running $\nsrun$ is shown as a function of the
    number of e-foldings $N_k$ before the end of inflation for several
    values of the potential width $f$ (note that larger $N_k$
    corresponds to smaller values of $k$ as in \refeqn{Nk}.
    The (light blue) filled region corresponds to the values of $N$
    consistent with the standard post-inflation cosmology for
    $\rhoRH > (\textrm{1 GeV})^4$.
    }
  \label{fig:nsrun}
\end{figure}

In general, $\ns$ is not constant: its variation can be characterized
by its running, $\nsrun$.  In this section, we use numerical solutions
to the equation of motion, \refeqn{eom}, as the slow roll approximation
(to the order used throughout this paper) is inaccurate for determining
the running.  As shown in \reffig{nsrun}, natural inflation predicts a
small, $\orderof{10^{-3}}$, negative spectral index running.  This is
negligibly small for WMAP sensitivities and this model is essentially
indistinguishable from zero running in the WMAP analysis.  While WMAP
data prefer a non-zero, negative running of $\orderof{10^{-1}}$ when
running is included in the analysis, zero running is not excluded at
95\% C.L. In Ref.~\cite{Easther:2006tv}, it was shown that the WMAP3
central value for the running would result in $N<30$ for single field,
slow roll inflation, an insufficient amount of expansion to solve the
cosmological problems for which inflation was proposed (\eg\ flatness of
the universe).  Reanalysis of the WMAP data with an $N>30$ prior removes
the preference for non-zero running and tightly constrains the running
to be, at most, of $\orderof{10^{-2}}$ \cite{Peiris:2006sj}; this
result, however, applies only for single field inflation models for
which the slow roll formalism is valid.  Analysis of WMAP data when
combined with Lyman-$\alpha$ forest and Supernovae Ia data also suggests
a smaller $\orderof{10^{-2}}$ running that is consistent with zero at
about the 1$\sigma$ level \cite{Seljak:2006bg}.

\begin{figure*}
  \insertfig{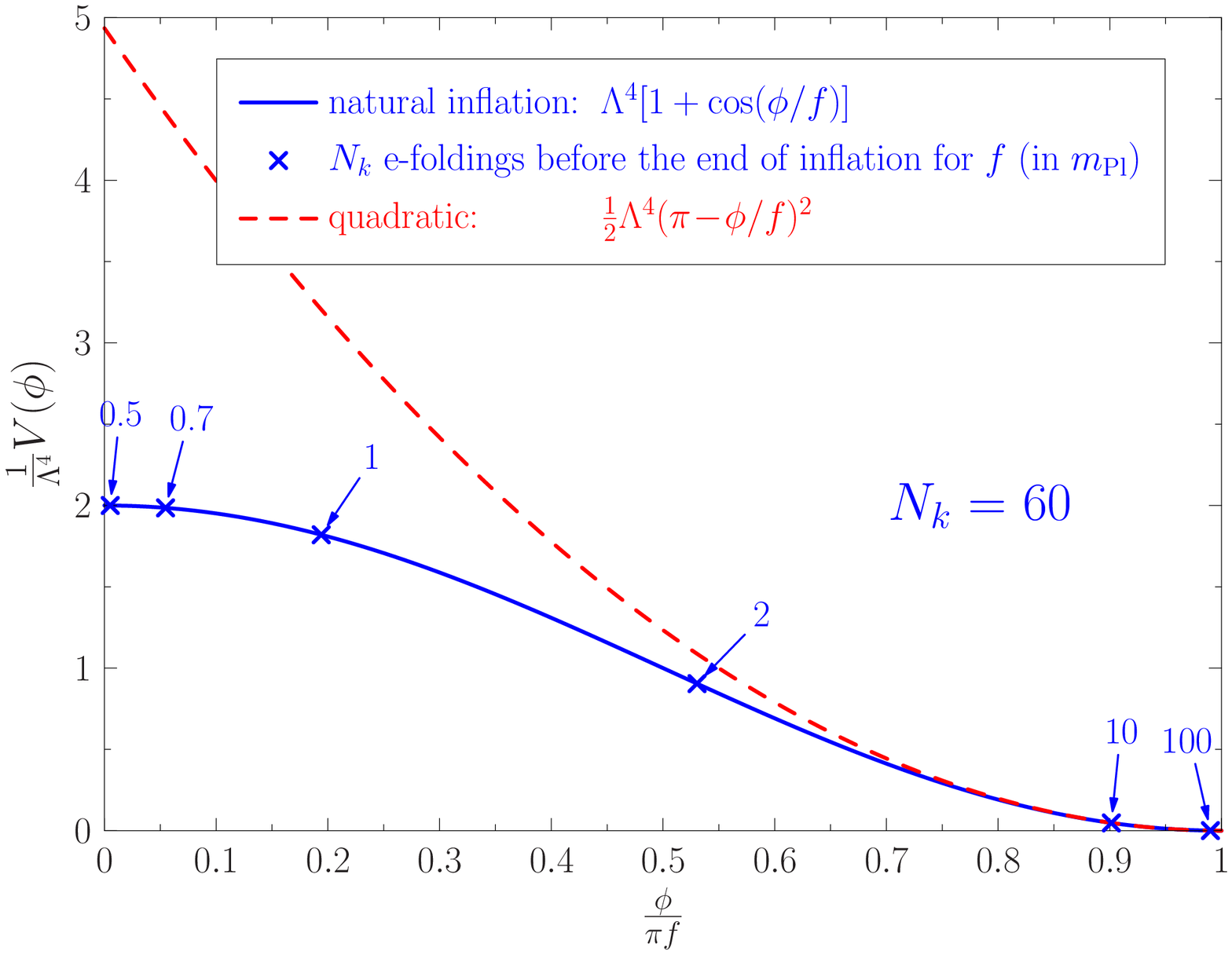}
  \hspace{\stretch{1}}
  \insertfig{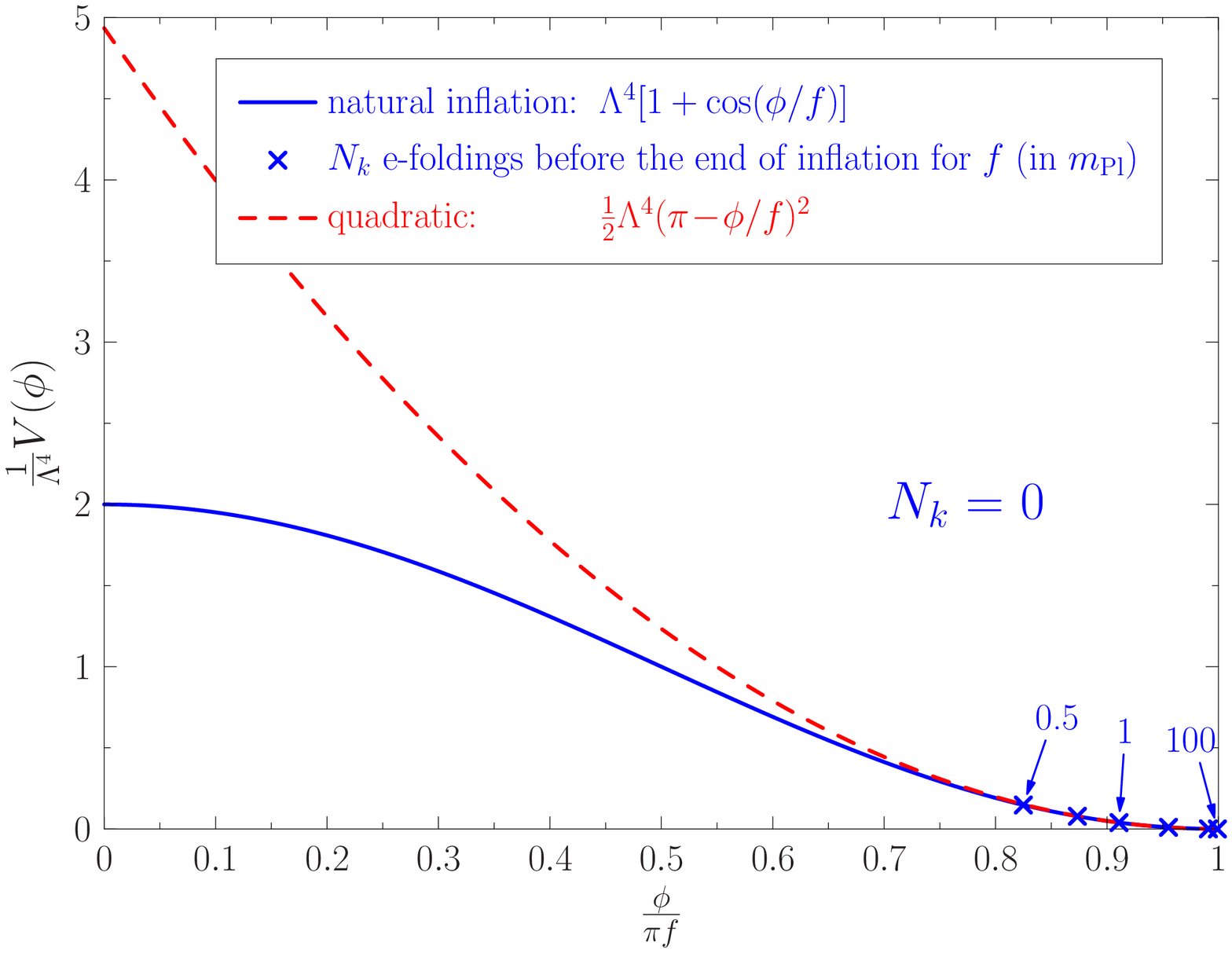}
  \caption[Inflaton potential]{
    The natural inflation potential is shown, along with a quadratic 
    expansion around the potential minimum.  Also shown are
    the positions on the potential at 60 e-foldings prior to the end
    of inflation (left panel) and at the end of inflation (right panel)
    for potential widths $f=(0.5,0.7,1,2,10,100) \mpl$.  For
    $f \gae 3\mpl$, the relevant portion of the potential is essentially
    quadratic during the last 60 e-foldings of inflation.
    }
  \label{fig:potential}
\end{figure*}

Small scale CMB experiments such as CBI \cite{Mason:2002tm}, ACBAR
\cite{Kuo:2002ua}, and VSA \cite{Dickinson:2004yr} will provide more
stringent tests of the running and hence of specific inflation models.
The predicted running for NI is too small to be detected in even these
experiments: if these experiments definitively detect a strong running
(\ie, excluding a zero/trivial running), natural inflation in the form
discussed here would be ruled out.

\section{\label{sec:Potential} Inflaton Potential and Inflationary Model
                Space}

In this section, we will examine the evolution of the inflaton field
$\phi$ along the potential.  We will show that the location on the
potential at which the final $\sim$60 e-foldings of inflation occurs
depends on the width $f$ of the potential.  We will also show that
natural inflation can fall into either the `large field' or `small
field' categorization defined by \cite{Dodelson:1997hr}, depending again
on the value of $f$.


The natural inflation potential is shown in
\reffig{potential}.  For comparison, a quadratic expansion
about the minimum at $\phi = \pi f$ is also shown.  Inflation occurs
when the field slowly rolls down the potential and ends at the point
where the field begins to move rapidly (technically, when
$\epsilon \ge 1$).  In the right panel of the figure, we show the
location along the potential where inflation ends ($N_k=0$) for various
values of the potential width $f$.  In the left panel, the location
along the potential is shown at $N_k=60$ e-foldings prior to the end of
inflation, the approximate time when fluctuations were produced that
correspond to the current horizon.  This is not necessarily where
inflation began: the field may have started at any point further up the
potential and produced more than 60 e-foldings of expansion.  The
rolling of the field above these points, however, would have produced
modes which are still on super-horizon scales today and hence are
unobservable.  In the following discussion, we will be referring only to
the \textit{observable} ($N_k \lae 60$) portion of the inflaton
evolution.  For all $f \gae 0.5\mpl$, inflation ends somewhere near the
bottom of the potential, with inflation for larger $f$ ending farther
down the potential than for smaller $f$.  We can see, however, that the
start of the observable portion of rolling is spread widely over the
potential.  For $f \lae 1\mpl$, current horizon modes were produced
while the field was near the top of the potential.  Conversely, for
$f \gae 3\mpl$, those modes were produced near the bottom of the
potential.  For $f \geq 5 \mpl$, the observationally relevant portion
of the potential is essentially a $\phi^2$ potential; note, however,
that in natural inflation this effectively power law potential is
produced via a natural mechanism.


\begin{figure}
  \insertfig{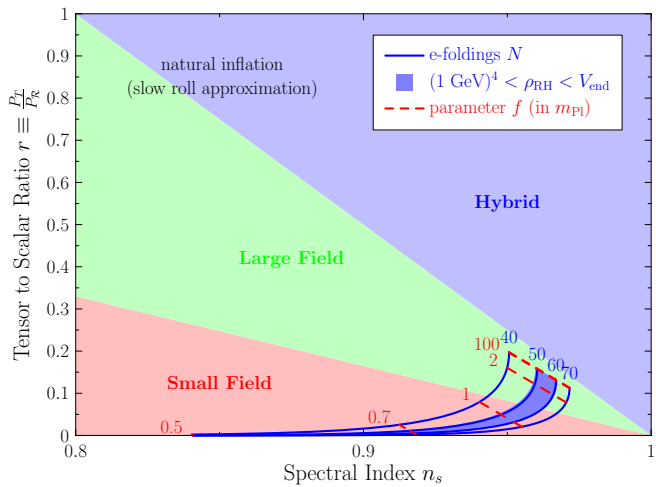}
  \caption[$r-n_s$ plane]{
    Natural inflation predictions in the $r$-$\ns$ plane (parameters
    and regions labeled as in \reffig{rnplane}), as well as the regions
    classifying small field, large field, and hybrid inflation models.  
    Natural inflation falls into different classes depending on the
    potential width $f$: for $f \lae 1.5\mpl$, natural inflation can be
    classified as a small field model, while for $f \gae 1.5\mpl$,
    natural inflation can be classified as a large field model.
    }
  \label{fig:rnplane2}
\end{figure}

Due to the variety of inflation models, there have been attempts to
classify models into a few groups.  Dodelson, Kinney \& Kolb
\cite{Dodelson:1997hr} have proposed a scheme with three categories:
small field, large field, and hybrid inflation models, which are easily
distinguishable in the SR approximation by the SR parameters $\epsilon$
and $\eta$.  Small field models are characterized by $\Vpp(\phi) < 0$
and $\eta < -\epsilon$, large field models by $\Vpp(\phi) > 0$ and
$-\epsilon < \eta \leq \epsilon$, and hybrid models by $\Vpp(\phi) > 0$
and $\eta > \epsilon >0$.  To first order in slow roll,
$\ns = 1-4\epsilon-2\eta$ and $r = 16\epsilon$, so the categories have
distinct regions in the $r$-$\ns$ plane, as shown in \reffig{rnplane2}.
Also shown in the figure are the predictions for natural inflation; 
parameters are labeled as in \reffig{rnplane} (which showed the same
predictions, albeit with a logarithmic rather than linear scale).  From
\reffig{rnplane2}, it can be seen that natural inflation does not fall
into a single category, but may be either small field or large field,
depending on the potential width $f$.  This should not be surprising
from the preceding discussion of the potential.  For $f \lae 1.5\mpl$,
$\phi$ is on the upper part of the potential, where $\Vpp(\phi) < 0$,
at $N_k=60$ and, thus, falls into the small field regime.  For
$f \gae 1.5\mpl$, $\phi$ is lower down the potential, where
$\Vpp(\phi) > 0$, at $N_k=60$ and falls into the large field regime
along with power law ($V(\phi) \sim \phi^p$ for $p>1$) models. (The 
large field regime for NI was first noted by Alabidi and Lyth in
Ref.~\cite{Alabidi:2005qi}.)  The WMAP3 constraints shown in
\reffig{rnplane} and discussed in \refsec{Fluctuations}, requiring
$f \gae 0.7\mpl$, still allow natural inflation to fall into either of
the small or large field categories.

\section{\label{sec:Conclusion} Conclusion}

Remarkable advances in cosmology have taken place in the past decade
thanks to Cosmic Microwave Background experiments.  The release of the
3 year data set by the Wilkinson Microwave Anisotropy Probe is leading
to exciting times for inflationary cosmology.  Not only are generic
predictions of inflation confirmed (though there are still outstanding
theoretical issues), but indeed individual inflation models are
beginning to be tested.

Currently the natural inflation model, which is extremely
well-motivated on theoretical grounds of naturalness, is a good fit to
existing data.  In this paper, we showed that for potential width $f >
0.7 \mpl$ and height $\Lambda \sim \mgut$ the model is in good
agreement with WMAP3 data.  Natural inflation predicts very little
running, an order of magnitude lower than the sensitivity of WMAP.
The location of the field in the potential while perturbations on
observable scales are produced was shown to depend on the width
$f$. Even for values $f>5 \mpl$ where the relevant parts of the
potential are indistinguishable from quadratic, natural inflation
provides a framework free of fine-tuning for the required potential.

There has been some confusion in the literature as to whether natural
inflation should be characterized as a `small-field' or `large-field'
model.  In \reffig{rnplane2} we demonstrated that either categorization
is possible, depending on the value of $f$, and that both are in
agreement with data.

Natural inflation makes definite predictions for tensor modes, as shown
in \reffig{rnplane}. Of particular significance is that current
observational constraints place a \textit{lower limit} on the
tensor/scalar ratio for Natural Inflation of order $10^{-3}$, a value
which is within range of proposed future high-precision cosmological
measurements \cite{Kinney:1998md,Friedman:2006zt}. Therefore Natural
Inflation represents a model which is both well-motivated and testable.


\begin{acknowledgments}
  CS and KF acknowledge the support of the DOE and the Michigan
  Center for Theoretical Physics via the University of Michigan.
  CS also acknowledges the support of the William I.\ Fine Theoretical
  Physics Institute at the University of Minnesota.
  WHK is supported in part by the National Science Foundation under
  grant NSF-PHY-0456777.
  KF thanks R.~Easther, M.~Turner, and L.~Verde for useful discussions.
\end{acknowledgments}





\begin{thebibliography}{99}

\bibitem{Guth:1980zm}
  A.~H.~Guth,
  Phys.\ Rev.\ D {\bf 23}, 347 (1981).

\bibitem{Kazanas:1980tx}
  D.~Kazanas,
  Astrophys.\ J.\  {\bf 241}, L59 (1980).

\bibitem{Starobinsky:1980te}
  A.~A.~Starobinsky,
  Phys.\ Lett.\ B {\bf 91}, 99 (1980).

\bibitem{Sato:1981ds}
  K.~Sato,
  Phys.\ Lett.\ B {\bf 99}, 66 (1981).

\bibitem{Sato:1980yn}
  K.~Sato,
  Mon.\ Not.\ Roy.\ Astron.\ Soc.\  {\bf 195}, 467 (1981).

\bibitem{Spergel:2006hy}
  D.~N.~Spergel {\it et al.},
  arXiv:astro-ph/0603449.

\bibitem{Alabidi:2006qa}
  L.~Alabidi and D.~H.~Lyth,
  arXiv:astro-ph/0603539.

\bibitem{Peiris:2006ug}
  H.~Peiris and R.~Easther,
  JCAP {\bf 0607}, 002 (2006)
  [arXiv:astro-ph/0603587].

\bibitem{Easther:2006tv}
  R.~Easther and H.~Peiris,
  JCAP {\bf 0609}, 010 (2006)
  [arXiv:astro-ph/0604214].

\bibitem{Seljak:2006bg}
  U.~Seljak, A.~Slosar and P.~McDonald,
  arXiv:astro-ph/0604335.

\bibitem{Kinney:2006qm}
  W.~H.~Kinney, E.~W.~Kolb, A.~Melchiorri and A.~Riotto,
  Phys.\ Rev.\ D {\bf 74}, 023502 (2006)
  [arXiv:astro-ph/0605338].

\bibitem{Martin:2006rs}
  J.~Martin and C.~Ringeval,
  JCAP {\bf 0608}, 009 (2006)
  [arXiv:astro-ph/0605367].

\bibitem{Peiris:2006sj}
  H.~Peiris and R.~Easther,
  arXiv:astro-ph/0609003.

\bibitem{Freese:1990rb}
  K.~Freese, J.~A.~Frieman and A.~V.~Olinto,
  Phys.\ Rev.\ Lett.\  {\bf 65}, 3233 (1990).

\bibitem{Freese:2004un}
  K.~Freese and W.~H.~Kinney,
  Phys.\ Rev.\ D {\bf 70}, 083512 (2004)
  [arXiv:hep-ph/0404012].

\bibitem{Spergel:2003cb}
  D.~N.~Spergel {\it et al.}  [WMAP Collaboration],
  Astrophys.\ J.\ Suppl.\  {\bf 148}, 175 (2003)
  [arXiv:astro-ph/0302209].

\bibitem{Adams:1992bn}
  F.~C.~Adams, J.~R.~Bond, K.~Freese, J.~A.~Frieman and A.~V.~Olinto,
  Phys.\ Rev.\ D {\bf 47}, 426 (1993)
  [arXiv:hep-ph/9207245].

\bibitem{Moroi:2000jr}
  T.~Moroi and T.~Takahashi,
  Phys.\ Lett.\ B {\bf 503}, 376 (2001)
  [arXiv:hep-ph/0010197].

\bibitem{Smoot:1992td}
  G.~F.~Smoot {\it et al.},
  Astrophys.\ J.\  {\bf 396}, L1 (1992).

\bibitem{Alabidi:2005qi}
  L.~Alabidi and D.~H.~Lyth,
  JCAP {\bf 0605}, 016 (2006)
  [arXiv:astro-ph/0510441].

\bibitem{Adams:1990pn}
  F.~C.~Adams, K.~Freese and A.~H.~Guth,
  Phys.\ Rev.\ D {\bf 43}, 965 (1991).

\bibitem{Bassett:2005xm}
  B.~A.~Bassett, S.~Tsujikawa and D.~Wands,
  Rev.\ Mod.\ Phys.\  {\bf 78}, 537 (2006)
  [arXiv:astro-ph/0507632].

\bibitem{Holman:1984yj}
  R.~Holman, P.~Ramond and G.~G.~Ross,
  Phys.\ Lett.\ B {\bf 137}, 343 (1984).

\bibitem{Kinney:1995xv}
  W.~H.~Kinney and K.~T.~Mahanthappa,
  Phys.\ Rev.\ D {\bf 52}, 5529 (1995)
  [arXiv:hep-ph/9503331].

\bibitem{Kinney:1995cc}
  W.~H.~Kinney and K.~T.~Mahanthappa,
  Phys.\ Rev.\ D {\bf 53}, 5455 (1996)
  [arXiv:hep-ph/9512241].

\bibitem{Kawasaki:2000yn}
  M.~Kawasaki, M.~Yamaguchi and T.~Yanagida,
  Phys.\ Rev.\ Lett.\  {\bf 85}, 3572 (2000)
  [arXiv:hep-ph/0004243].

\bibitem{Arkani-Hamed:2003wu}
  N.~Arkani-Hamed, H.~C.~Cheng, P.~Creminelli and L.~Randall,
  Phys.\ Rev.\ Lett.\  {\bf 90}, 221302 (2003)
  [arXiv:hep-th/0301218].

\bibitem{Arkani-Hamed:2003mz}
  N.~Arkani-Hamed, H.~C.~Cheng, P.~Creminelli and L.~Randall,
  JCAP {\bf 0307}, 003 (2003)
  [arXiv:hep-th/0302034].

\bibitem{Kaplan:2003aj}
  D.~E.~Kaplan and N.~J.~Weiner,
  JCAP {\bf 0402}, 005 (2004)
  [arXiv:hep-ph/0302014].

\bibitem{Firouzjahi:2003zy}
  H.~Firouzjahi and S.~H.~H.~Tye,
  Phys.\ Lett.\ B {\bf 584}, 147 (2004)
  [arXiv:hep-th/0312020].

\bibitem{Hsu:2004hi}
  J.~P.~Hsu and R.~Kallosh,
  JHEP {\bf 0404}, 042 (2004)
  [arXiv:hep-th/0402047].

\bibitem{Freese:1994fp}
  K.~Freese,
  Phys.\ Rev.\ D {\bf 50}, 7731 (1994)
  [arXiv:astro-ph/9405045].

\bibitem{Adams:1991ma}
  F.~C.~Adams and K.~Freese,
  Phys.\ Rev.\ D {\bf 43}, 353 (1991)
  [arXiv:hep-ph/0504135].

\bibitem{Kim:2004rp}
  J.~E.~Kim, H.~P.~Nilles and M.~Peloso,
  JCAP {\bf 0501}, 005 (2005)
  [arXiv:hep-ph/0409138].

\bibitem{Dimopoulos:2005ac}
  S.~Dimopoulos, S.~Kachru, J.~McGreevy and J.~G.~Wacker,
  arXiv:hep-th/0507205.

\bibitem{Lidsey:1995np}
  J.~E.~Lidsey, A.~R.~Liddle, E.~W.~Kolb, E.~J.~Copeland, T.~Barreiro
  and M.~Abney,
  Rev.\ Mod.\ Phys.\  {\bf 69}, 373 (1997)
  [arXiv:astro-ph/9508078].

\bibitem{Liddle:2003as}
  A.~R.~Liddle and S.~M.~Leach,
  Phys.\ Rev.\ D {\bf 68}, 103503 (2003)
  [arXiv:astro-ph/0305263].

\bibitem{Guth:1982ec}
  A.~H.~Guth and S.~Y.~Pi,
  Phys.\ Rev.\ Lett.\  {\bf 49}, 1110 (1982).

\bibitem{Hawking:1982cz}
  S.~W.~Hawking,
  Phys.\ Lett.\ B {\bf 115}, 295 (1982).

\bibitem{Starobinsky:1982ee}
  A.~A.~Starobinsky,
  Phys.\ Lett.\ B {\bf 117}, 175 (1982).

\bibitem{Mukhanov:1985rz}
  V.~F.~Mukhanov,
  JETP Lett.\  {\bf 41}, 493 (1985)
  [Pisma Zh.\ Eksp.\ Teor.\ Fiz.\  {\bf 41}, 402 (1985)].

\bibitem{Mukhanov:1988jd}
  V.~F.~Mukhanov,
  Sov.\ Phys.\ JETP {\bf 67}, 1297 (1988)
  [Zh.\ Eksp.\ Teor.\ Fiz.\  {\bf 94N7}, 1 (1988)].

\bibitem{Mukhanov:1990me}
  V.~F.~Mukhanov, H.~A.~Feldman and R.~H.~Brandenberger,
  Phys.\ Rept.\  {\bf 215}, 203 (1992).

\bibitem{Stewart:1993bc}
  E.~D.~Stewart and D.~H.~Lyth,
  Phys.\ Lett.\ B {\bf 302}, 171 (1993)
  [arXiv:gr-qc/9302019].

\bibitem{Dodelson:1997hr}
  S.~Dodelson, W.~H.~Kinney and E.~W.~Kolb,
  Phys.\ Rev.\ D {\bf 56}, 3207 (1997)
  [arXiv:astro-ph/9702166].

\bibitem{Kinney:1998md}
  W.~H.~Kinney,
  Phys.\ Rev.\ D {\bf 58}, 123506 (1998)
  [arXiv:astro-ph/9806259].

\bibitem{winstein}
  See \eg\ B.~Winstein,
  2nd Irvine Cosmology Conference
  (2006).

\bibitem{unknown:2006uk}
    [Planck Collaboration],
  arXiv:astro-ph/0604069.

\bibitem{Bock:2006yf}
  J.~Bock {\it et al.},
  arXiv:astro-ph/0604101.

\bibitem{Mason:2002tm}
  B.~S.~Mason {\it et al.},
  Astrophys.\ J.\  {\bf 591}, 540 (2003)
  [arXiv:astro-ph/0205384].

\bibitem{Kuo:2002ua}
  C.~l.~Kuo {\it et al.}  [ACBAR collaboration],
  Astrophys.\ J.\  {\bf 600}, 32 (2004)
  [arXiv:astro-ph/0212289].

\bibitem{Dickinson:2004yr}
  C.~Dickinson {\it et al.},
  Mon.\ Not.\ Roy.\ Astron.\ Soc.\  {\bf 353}, 732 (2004)
  [arXiv:astro-ph/0402498].

\bibitem{Friedman:2006zt}
  B.~C.~Friedman, A.~Cooray and A.~Melchiorri,
  arXiv:astro-ph/0610220.


\end{thebibliography}
\end{document}